# Self-assembled Nanocapsules in Water: A Molecular Mechanism Study

Hang Xiao [a,b], Xiaoyang Shi [a] and Xi Chen [a,b] *

[a] *Columbia Nanomechanics Research Center, Department of Earth and Environmental Engineering, Columbia University, New York, NY 10027, USA*

[b] *School of Chemical Engineering, Northwest University, Xi'an 710069, China*

## Abstract

The self-assembly mechanism of one-end-open carbon nanotubes (CNTs) suspended in an aqueous solution was studied by molecular dynamics simulations. It was shown that two one-end-open CNTs with different diameters can coaxially self-assemble into a nanocapsule. The nanocapsules formed were stable in aqueous solution under ambient conditions, and the pressure inside the nanocapsule was much higher than the ambient pressure due to the van der Waals interactions between two parts of the nanocapsule. The effect of the normalized radius difference, normalized inter-tube distance and aspect ratio of the CNT pairs were systematically explored. The electric field response of nanocapsules was studied with *ab initio* molecular dynamics simulations, which shows that nanocapsules can be opened by applying an external electric field, due to the polarization of carbon atoms. This discovery not only sheds light on a simple yet robust nanocapsule self-assembly mechanism, but also underpins potential innovations in drug delivery, nano-reactors, etc.

* Corresponding author: Department of Earth and Environmental Engineering, Columbia University, New York, NY 10027, USA. E-mail: xichen@columbia.edu

# 1. Introduction

Micro- and nano-scale capsules are of great interest due to their potential applications in many fields, including drug delivery, adsorbents, nano-reactors, to name a few. The polymer-based nanocapsules has been extensively studied for drug delivery in the pharmaceutical field [1–6]. The protective coating in these nanocapsules is usually pyrophoric and easily oxidized, to release the therapeutic substance confined inside [6]. Substances confined within nanoscale space may exhibit unique physical and chemical properties. Giovambattista et al [7] studied the nanoconfinement induced phase transitions in liquid water. Shi et al [8] investigated the unconventional reversible chemical reaction driven by nanoconfined ion hydration. The nanoconfined space and pressure provided by a nanocapsule enable its potential application as nano-reactor.

Carbon nanotubes (CNTs) are cylindrical forms of graphene layers with either open or close ends [9,10]. Their outstanding electrical and thermal conductivity, and superior strength-to-density and stiffness-to-density ratios have stimulated increasing interests [11–15]. Nanocapsules self-assembled by CNTs can be ideal vehicles for drug delivery, since CNTs are non-immunogenic [16] and can be functionalized [17–20].

Herein, we study the self-assembly of one-end-open CNTs into nanocapsules in water, showing that two one-end-open CNTs with different diameters, can coaxially self-assemble into a nanocapsule that is stable in water under ambient conditions. The effect of the normalized radius difference, normalized inter-tube distance, aspect ratio of the CNT pairs are systematically studied. The effect of electric field on the structure of nanocapsules is investigated with *ab initio* molecular dynamics (AIMD) simulations, showing that nanocapsules can be disassembled by applying an external electric field. This discovery not only reveals a simple yet robust nanocapsule self-assembly mechanism, but also sheds light on the potential applications in drug delivery, nano-reactors, etc.

## 2. Model and Method

Molecular dynamics (MD) calculations were performed using LAMMPS code [21], for the self-assembly of one-end-open CNTs into nanocapsules in an orthorhombic water box under ambient temperature (T = 300 K) and pressure (P = 1 bar). The straight part of the one-end-open CNT was described by Morse bonds, harmonic valence angles, harmonic torsion angles and Lennard-Jones (LJ) 12-6 pair interactions [22]. The cap of the one-end-open CNT was fixed rigid, since its deformation during the self-assembly process was negligible. Water molecules were modeled by the TIP3P-ew model [23] and the long range electrostatic interactions were calculated using the PPPM algorithm [24]. Following Hummer et al. [25], the interactions between CNTs and water molecules were described by a LJ-potential between oxygen and carbon. The equation of motion was solved with a velocity Verlet algorithm, using a time step of 1.0 fs, which led to stable dynamics trajectories.

A pair of one-end-open CNTs were initially coaxially aligned (constrained) with their open-end facing each other (see Fig. 1). The initial constrained distance between the open ends of two CNTs was 2 Å. The system with constrained CNTs in water was first equilibrated at 300 K and 1 atmospheric pressure with the NPT (constant number of particles, constant pressure and constant temperature) ensemble for 500 ps. Constraints on CNTs were removed after the equilibration step such that they were free to move during the self-assembly process.

To study the effect of electric field on the structure of nanocapsules, *ab initio* molecular dynamics simulations (shown in Fig. 8) were performed at the PBE [26] /GTH-DZVP [27] level in the NVT (constant number of particles, constant volume and constant temperature) ensemble of the CP2K [28] code. The empirical dispersion correction schemes proposed by Grimme (D3) [29] was used in combination with PBE functional to account for the van der Waals interactions. The external electric field was applied along the axis of nanocapsule. Water solvent outside the nanocapsule was not considered in *ab initio* molecular dynamics simulations, due to its negligible effect on the electric field response of the nanocapsule.

# 3. Results and Discussions

The self-assembly process of the nanocapsule by one-end-open (8, 8) and (13, 13) CNTs is shown in Fig. 1. After the equilibration step mentioned in section 2, the CNTs were fully solvated (see Fig. 1(A)). Once the constraints on CNTs were removed, the smaller tube gradually found its way into the larger tube, forming a stable nanocapsule. The self-assembly process was roughly comprised of 3 steps: (1) Two tubes came close to each other, due to inter-tube vdW attractions (Fig. 1(a-b)). (2) Excessive water molecules were discharged through the opening formed by the rotation of tubes (Fig. 1(b-d)). (3) Two tubes became coaxially aligned and the smaller tube was partially inserted into the larger counterpart (Fig. 1(d-e)). In Fig. 2, it shows the center-of-mass (COM) distance and the vdW interaction energy between two CNTs as a function of time. The three steps of self-assembly process can be clearly identified in Fig.2: (1) the COM distance between two CNTs decreased rapidly in the first 15 ps; (2) the COM distance between two CNTs remained more or less the same during the second step; (3) the huge reduction of the COM distance between two CNTs indicates the quick insertion process. The magnitude of inter-tube vdW interaction energy increased as the inter-tube COM distance decreased, confirming that the vdW interaction is the driving force of the nanocapsule self-assembly.

## 3.1 Effect of normalized radius difference $\Delta R/r_m$

The self-assembly process is expected to be strongly dependent on the inter-tube spacing in the radial direction, since the driving force of the self-assembly is the inter-tube vdW interaction. The inter-tube spacing is characterized by normalized radius difference, denoted as $\Delta R/r_m$, where $\Delta R$ is the difference of the radius of the two CNTs; $r_m$ is the distance at which the carbon-carbon LJ potential reaches its minimum ($r_m$ = 3.81 Å)). The nanocapsule self-assembly processes with different $\Delta R/r_m$ are shown in Fig. 3, which can be divided into three categories: (1) When $\Delta R/r_m$ was close to 1 ($\Delta R/r_m$= 0.89, 0.92 and 1.06), nanocapsules were successfully assembled. (2) When $\Delta R/r_m$ was relatively large ($\Delta R/r_m$= 1.24), a nanocapsule with less solvent trapped inside was assembled. The solvent escaped from the tubes during the self-assembly process due to large inter-tube spacing. (3) When $\Delta R/r_m$ was fairly large ($\Delta R/r_m = 1.42$) or too small ($\Delta R/r_m = 0.71$), the CNTs failed to form nanocapsules. For $\Delta R/r_m = 1.42$, the self-assembly failed due to weak inter-tube attraction. For $\Delta R/r_m = 0.71$, small inter-tube spacing rendered the nanocapsule self-assembly energetically unfavorable.

For $\Delta R/r_m = 0.92$ and $\Delta R/r_m = 0.89$, nanocapsules were self-assembled by zigzag and armchair CNT pairs, respectively. Nanocapsules with similar structures were self-assembled following the aforementioned 3-step route, indicating that the effect of chirality on the assembly process is negligible.

## 3.2 Effect of the normalized inter-tube distance $D/r_m$

Similarly, the initial inter-tube distance is expected to strongly affect the self-assembly of nanocapsules. The normalized inter-tube distance is defined as $D/r_m$, where $D$ is the initial axial distance of the open-ends of CNTs. The time evolution of nanocapsule self-assembly by one-end-open (8, 8) and (13, 13) CNTs with different $D/r_m$ are shown in Fig. 4. When the inter-tube distance was relatively small ($D/r_m \leq 1.31$), nanocapsules were successfully self-assembled, due to the relatively strong inter-tube vdW attractions. The self-assembly failed when the inter-tube distance is large ($D/r_m > 1.31$), due to weak inter-tube vdW attraction.

## 3.3 Effect of temperature

Fig. 5 shows the nanocapsule self-assembly map as both $D/r_m$ and $\Delta R/r_m$ are varied. The cases when the nanocapsule was assembled or not at 300 K and 350 K are separated by the solid red line and the dashed red line, respectively. The effect of temperature on the nanocapsule self-assembly can be evaluated by the temperature-induced shift of the parameter space boundary that distinguishes the successful self-assembly and failed ones. The dividing line shifts left as temperature rises, indicating that the nanocapsule self-assembly was not favored at relatively high temperature, due to the strong thermal fluctuations.

## 3.4 Effect of the aspect ratio $l/d$

The role of aspect ratio of CNT pairs is investigated by comparing the self-assembly of nanocapsules by one-end-open (8, 8) and (13, 13) CNTs with different $l/d$, where $l$ is the total length of the two CNTs in a nanocapsule (before assembly); $d$ is the average diameter of the CNTs. In Fig. 6, it shows that it took much less time for systems with large aspect ratios ($l/d$= 6.3 & 5.2) to form the nanocapsule than the systems with small aspect ratios (($l/d$= 2.5 & 3.8)). When $l/d$= 2.5 & 3.8, the small CNT rotated during the insertion process, prolonging the self-assembly process. On the contrary, an extremely fast insertion of the small tube into the large tube without perceptible rotation was observed when $l/d$= 6.3 & 5.2, since the longer tubes were more resistant to rotation in water.

Owing to the inter-tube vdW interaction, the pressure of water solvent inside the nanocapsule is expected to be higher than that of the solvent outside. Fig. 7(a) shows that both the pressure inside the nanocapsule and the magnitude of inter-tube interaction energy increased with increasing $l/d$. The structure of water inside the nanocapsule with $l/d = 5.2$ before and after the formation of the nanocapsule are shown in Fig. 7(b) and Fig. 7(c), respectively. The pressure inside the nanocapsule was on the order of 1 GPa. Such a high pressure triggered the formation of square ice inside the small tube with strong nano-confinement, as shown in Fig. 7(c). The square ice crystal formed in the nanocapsule is similar to the square ice structures formed in graphene nanocapillaries [30]. Recently, Vasu et al [31] reported the vdW pressure formed between graphene layers (on the order of 1 GPa) is able to induce chemical reactions of the trapped interlayer molecules. Therefore, the vdW pressure inside the nano-confined space of nanocapsule sheds light on its potential applications as nano-reactors.

It should be noted that in all simulations presented herein, the open ends of the two CNTs were in close proximity and aligned in the beginning of assembly process. If the two open ends were not aligned coaxially but they are still close to each other, with the aid of thermal fluctuation, the CNTs could still self-assembly into a nanocapsule although it takes a longer time. However, if the two CNTs were initially placed far away from each other, it would become difficult for them to attract to each other and assemble. Therefore, in practice, if one randomly places CNTs in a solution, the yield ratio of assembled nanocapsules may be low and depend on the relative density of CNTs. It is known that CNTs in solutions can be effectively aligned with an moderate external electric field, due to the electronic polarization [32–34]. Therefore, the yield ratio of nanocapsule self-assembly could be increased by applying a moderate external electric field.

### 3.5 Open the nanocapsule by an external electric field

Once the nanocapsule is self-assembled, the structural stability of the nanocapsule is maintained by the inter-tube vdW interaction energy. Therefore, controllable opening and closing of the nanocapsule can be achieved by manipulating the inter-tube interaction energy. The polarization of carbon atoms in CNTs can be triggered by an external electric field. Consequently, the inter-tube interaction can be controlled. For instance, it is shown that the opening and closing of the carbon nanoscrolls can be controlled by an electric field, due to polarization-induced change of surface adhesion [35]. Zhu et al [36] revealed that an external electric field can effectively tune the morphology of graphene nanocages, owing to the polarization of the carbon atoms.

The electric field response of the nanocapsule when the electric field intensity, $E$=0.25 V/Å and $E$=0.75 V/Å were studied with AIMD simulations. Time evolution of the morphology of nanocapsules with corresponding electrostatic potential maps under electric field are displayed in Fig. 8. Stronger polarization was observed in the nanocapsule under the electric field with $E$=0.75 V/Å, compared to the scenario with $E$=0.25 V/Å. While the relatively weak polarization induced by the electric field with $E$=0.25 V/Å was not able to open the nanocapsule, the nanocapsule was opened under $E$=0.75 V/Å in less than 1 ps. Our result show that an external electric field can reduce the inter-tube adhesion in the nanocapsule, enabling facile control of the nanocapsule morphology by tuning the external electric field. Owing to its non-immunogenic nature, chemical tunability via functionalization and electric-field controlled morphology, nanocapsules self-assembled by one-end-open CNTs can be ideal vehicles for drug delivery,

# 4. Concluding Remarks

In summary, molecular dynamics simulations show that one-end-open CNTs pairs with proper radius difference can coaxially self-assemble into a nanocapsule. The nanocapsules formed are stable in aqueous solution in ambient conditions and the pressure inside the nanocapsule is much higher than the pressure in the aqueous solution, due to the vdW attractions between the CNT pairs. The effect of the normalized radius difference, normalized inter-tube distance and aspect ratio of the CNT pairs were systematically explored. AIMD simulations show that nanocapsules can be opened by applying an external electric field, due to the polarization of the CNT pairs. Our results have general implications on fabricating nanocapsules with various building blocks such as nanotubes (e.g. open-ended CNTs), nano-bowls (e.g. $C_{50}H_{10}$ fullerene bowls), etc. In addition, the nanocapsules can be opened via external electric field, which sheds light on their potential applications in drug-delivery, nano-reactors, etc.

# Acknowledgments

X. C. acknowledges support from National Natural Science Foundation of China (11372241 and 11572238), Advanced Research Projects Agency-Energy (DE-AR0000396) and Air Force Office of Scientific Research (FA9550-12-1-0159).

# Figures

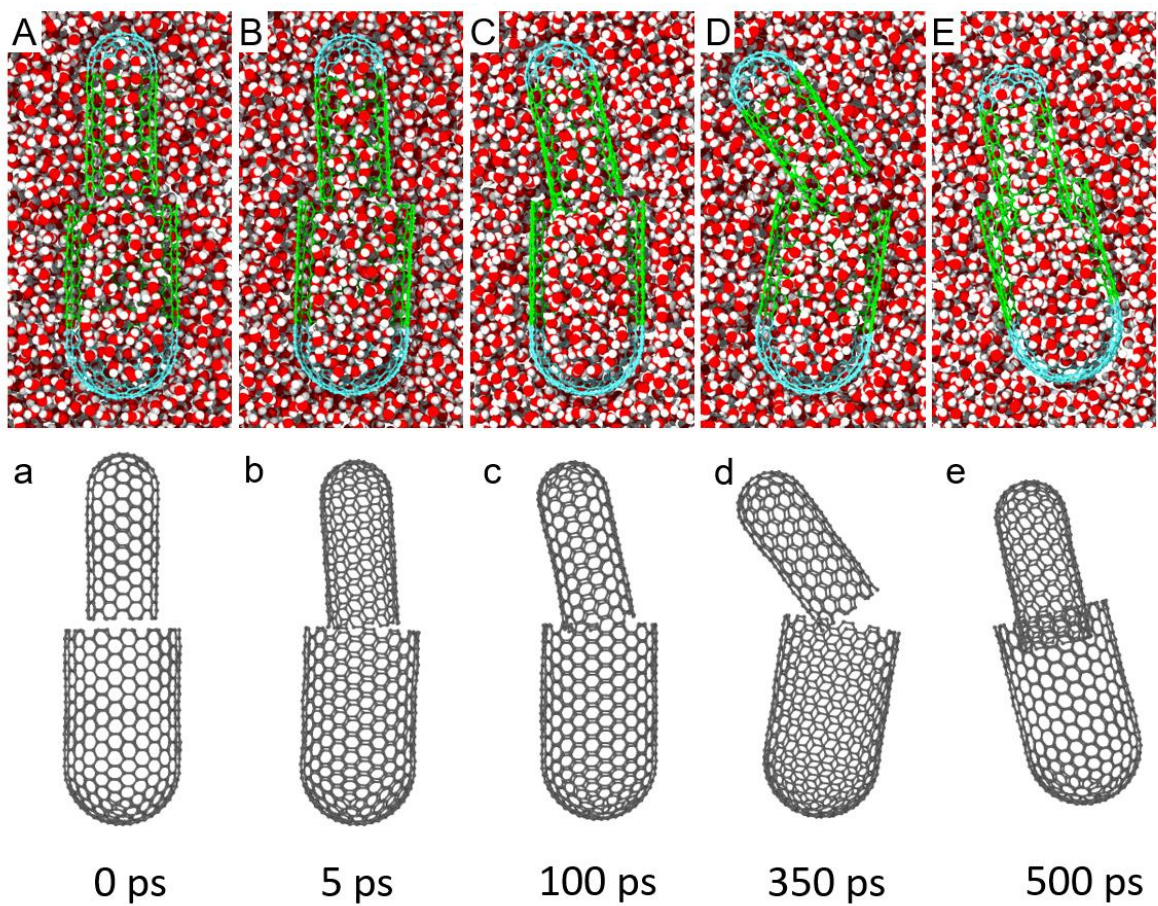

Fig. 1. Snapshots of the self-assembly process of the nanocapsule from one-end-open (8,8) and (13,13) CNTs. In A, B, C, D and E, nanocapsules are sliced in half to show the water molecules inside. The rigid caps (A, B,

C, D, E) are marked in cyan, and the straight regions described by the Morse bond model are marked in green. While in a, b, c, d and e, one-end-open CNTs are marked in grey and water molecules are not displayed for clarity. The small one-end-open CNT gradually finds its way into the large one-end-open CNT, forming a stable nanocapsule.

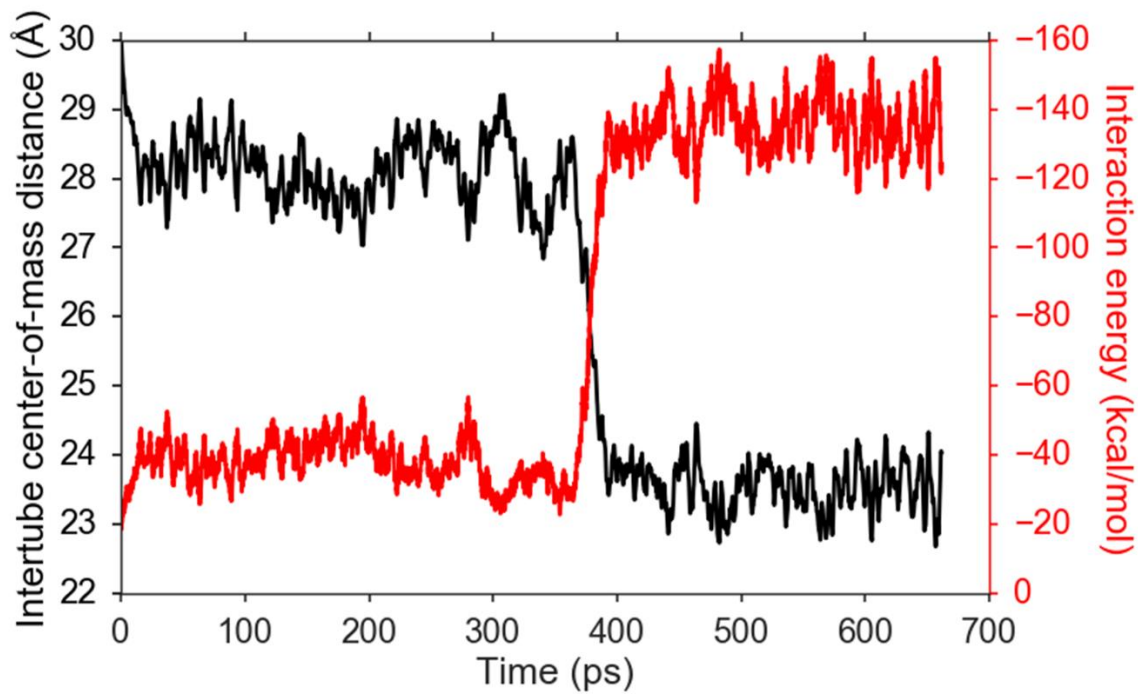

Fig. 2. The center-of-mass distance (in black) and the interaction energy (in red) between one-end-open (8,8) and (13,13) nanotubes as a function of time. The nanocapsule is formed at around 400 ps.

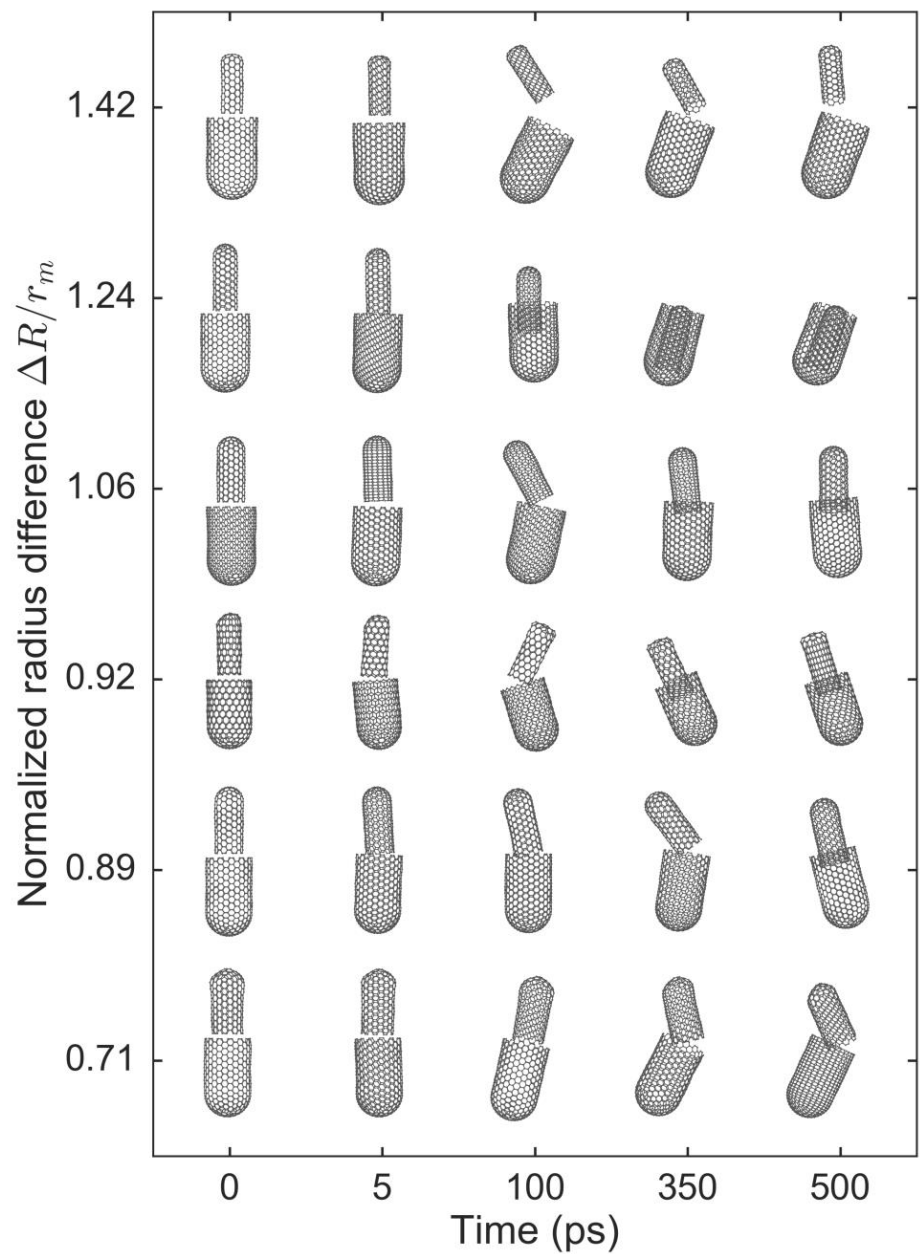

Fig. 3. Time evolution of the nanocapsule self-assembly by one-end-open CNT pairs with different normalized radius differences ($\Delta R/r_m$).

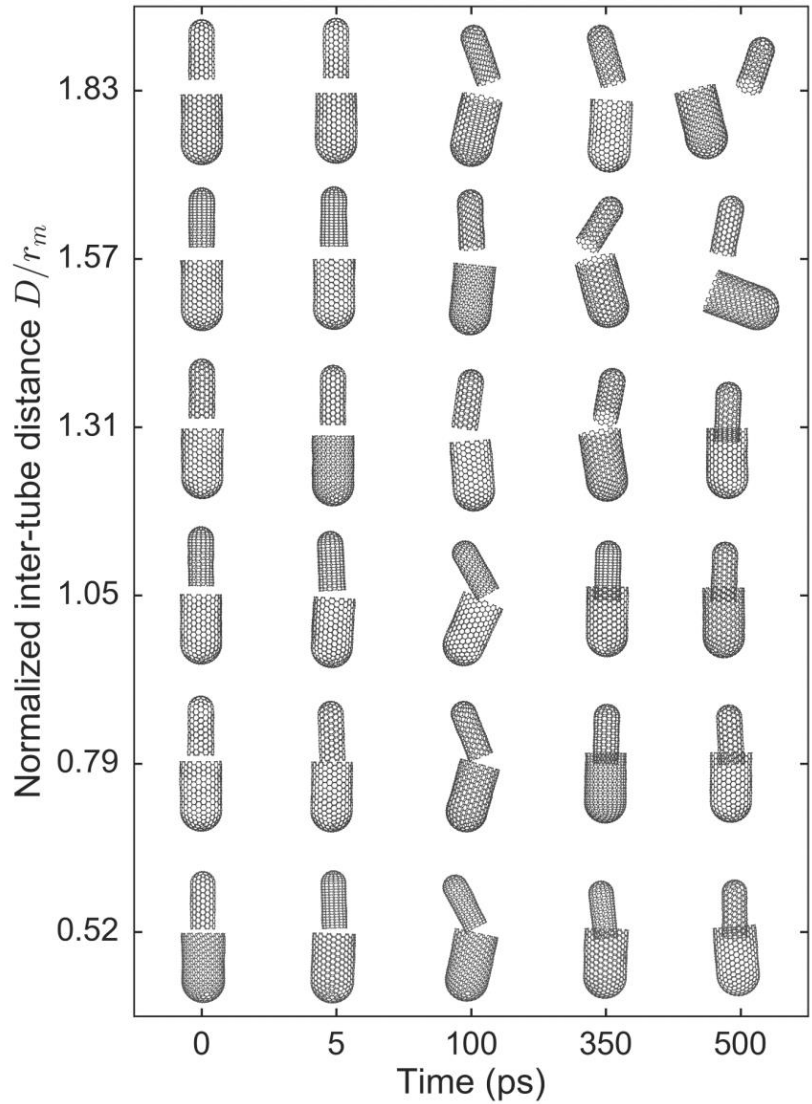

Fig. 4. Time evolution of the nanocapsule self-assembly by one-end-open CNT pairs ($\Delta R/r_m = 1.06$) with varying normalized inter-tube distances ($D/r_m$).

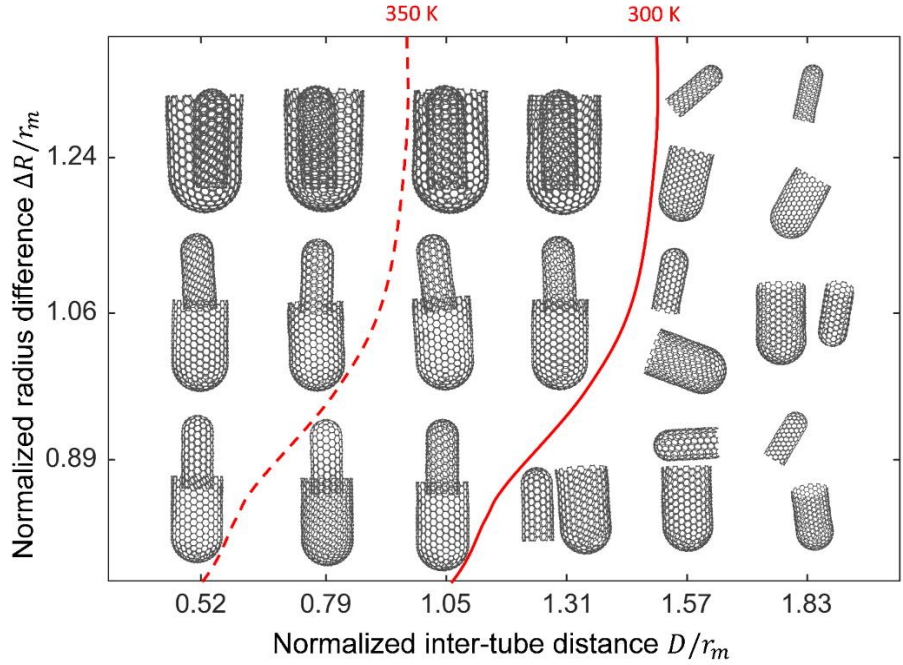

Fig. 5. The nanocapsule self-assembly map as both normalized inter-tube distance ($D/r_m$) and normalized radius difference ($\Delta R/r_m$) are varied. Snapshots of systems at time t=500 ps are shown. The cases when the nanocapsule is assembled or not at 300 K are separated by the solid red line. The scenarios when the nanocapsule is assembled or not at 350 K are separated by the dashed red line (the systems at 350 K are not shown).

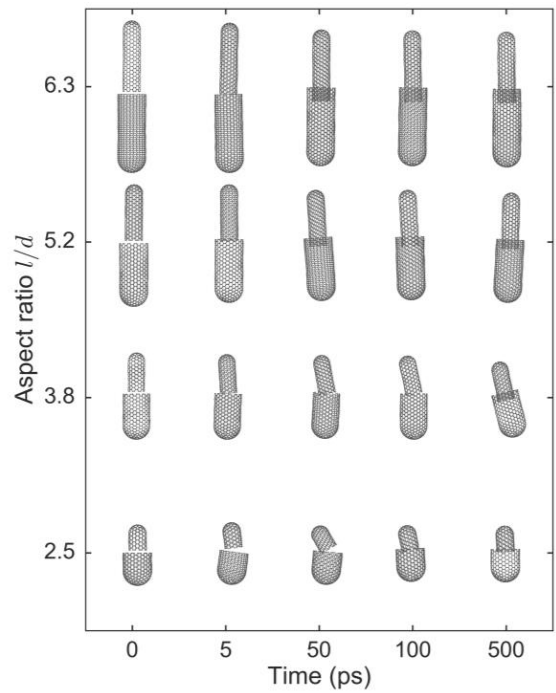

Fig. 6. Time evolution of the nanocapsule self-assembly by one-end-open CNT pairs ($\Delta R/r_m = 1.06$, $D/r_m = 0.52$) with different aspect ratio, $l/d$.

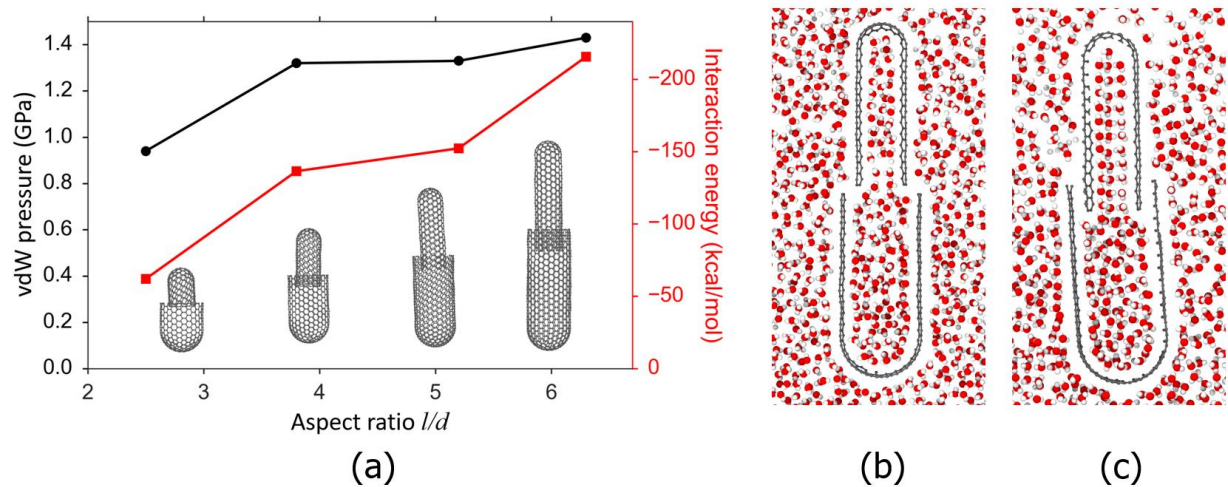

Fig. 7. (a) The van der Waals pressure inside the nanocapsule and the interaction energy between two CNTs as a function of the aspect ratio, *l/d*. Comparison of water structure in the CNTs before (b) and after (c) the nanocapsule is formed. Square ice structure is formed in the smaller CNT due to the high van der Waals pressure and the nano-confinement.

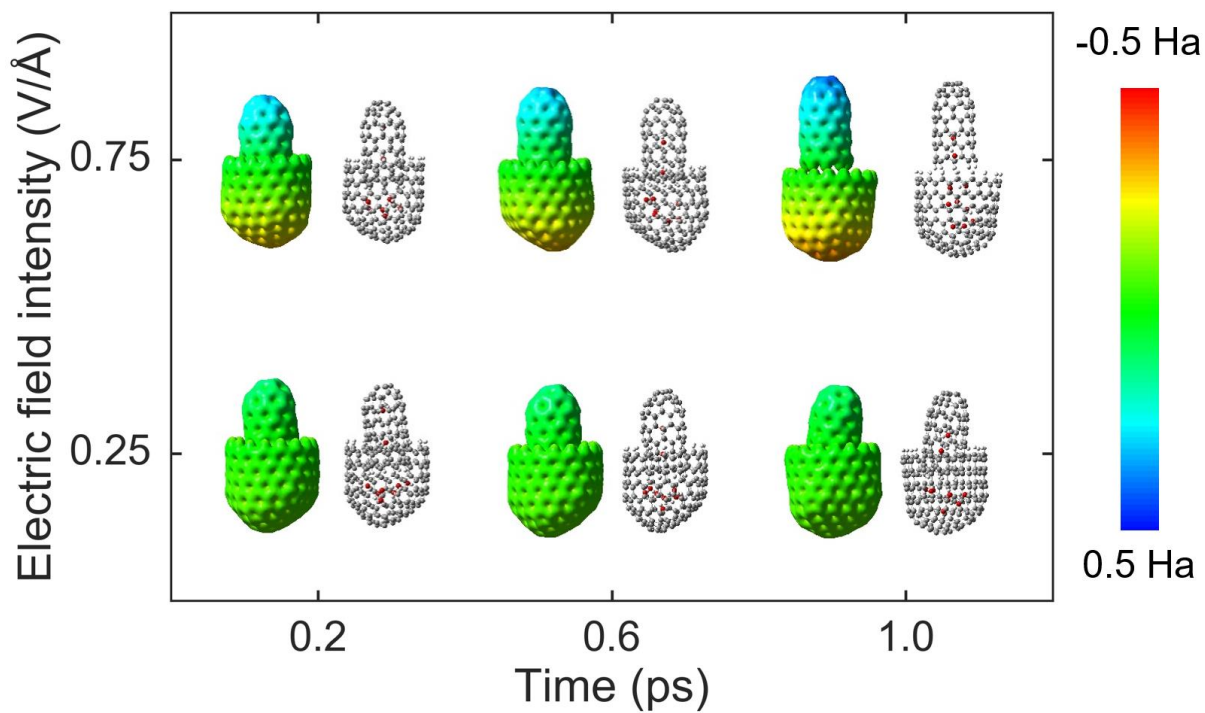

Fig. 8. The electric field response of the nanocapsule when $E$=0.25 V/Å and $E$=0.75 V/Å. The electrostatic potential (ESP) maps and the structures of the nanocapsule at t=0.2 ps, 0.6 ps and 1 ps are shown.